\documentclass[twocolumn,showpacs,prl,amsmath,amssymb,nobalancelastpage]{revtex4}
 \def\tskip{\setlength{\tskip}{5pt}}
\def\colwidth{\setlength{\colwidth}{3.5in}}

\usepackage{graphicx} \usepackage{dcolumn} \usepackage{bm}
\usepackage{epsfig} \usepackage{amssymb} \usepackage{amsmath}

\def\fun#1#2{\lower3.6pt\vbox{\baselineskip0pt\lineskip.9pt
\ialign{$\mathsurround=0pt#1\hfil##\hfil$\crcr#2\crcr\sim\crcr}}}

\begin{document}

\title{A new cosmic microwave background constraint to primordial gravitational waves}

\author{Tristan L. Smith, Elena Pierpaoli, and Marc
     Kamionkowski} \affiliation{California Institute of
     Technology, Mail Code 130-33, Pasadena, CA 91125}

\date{\today}
   
\begin{abstract}
Primordial gravitational waves (GWs) with frequencies
$\gtrsim10^{-15}$ Hz contribute to the radiation density
of the Universe at the time of decoupling of the cosmic microwave
background (CMB).  The effects of this GW background on the CMB
and matter power spectra are identical to those due to
massless neutrinos, unless the initial density-perturbation
amplitude for the gravitational-wave gas is non-adiabatic, as
may occur if such GWs are produced
during inflation or some post-inflation phase transition.  In
either case, current observations provide a constraint to the
GW amplitude that competes with that from 
big-bang nucleosynthesis (BBN), although it extends to much
lower frequencies ($\sim10^{-15}$ Hz rather than the
$\sim10^{-10}$ Hz lower limit from BBN): at 95\% confidence-level,
 $\Omega_{\mathrm{gw}}h^2 \lesssim 6.9\times
10^{-6}$ for homogeneous (i.e., non-adiabatic) initial conditions.  Future CMB
experiments, like Planck and CMBPol, should allow sensitivities to
$\Omega_{\mathrm{gw}}h^2 \lesssim  1.4 \times 10^{-6}$ and 
$\Omega_{\mathrm{gw}}h^2 \lesssim 5 \times10^{-7}$, respectively. 
\end{abstract}


\maketitle

There are many conjectured sources of a primordial cosmological
gravitational-wave background (CGWB), including inflation,
pre-big bang theories, phase transitions, or the ekpyrotic
model \cite{Maggiore00}.  Such backgrounds are among the targets of the Laser
Interferometric Gravitational-Wave Observatory (LIGO), and they
will be sought with future observatories, such as NASA's Laser
Interferometer Space Antenna (LISA), the Big Bang Observer
(BBO), and Japan's Deci-Hertz Interferometer Gravitational-wave
Observatory (DECIGO).

The CGWB amplitude is constrained at the lowest observable
frequencies, $\sim10^{-17}-10^{-16}$ Hz
(corresponding to wavelengths comparable to the cosmological
horizon today), by large-angle fluctuations in the cosmic
microwave background (CMB)
temperature \cite{first_papers}.
Prospects for probing lower CGWB
amplitudes at these frequencies come from future measurements of
the polarization of the CMB \cite{Kamionkowski:1996zd,Seljak:1996gy}.
Apart from a window around $10^{-9}-10^{-8}$ Hz, where the
CGWB is constrained by pulsar timing
\cite{Kaspi:1994hp, Lommen:2002je}, the strongest constraint to the CGWB
amplitude for frequencies greater than
$\sim10^{-10}$ Hz comes from big-bang nucleosynthesis (BBN)
\cite{Allen:1997}.  The lower limit to the frequency range is
determined by the comoving horizon size at the time of BBN.
Primordial gravitational waves of shorter wavelengths, or larger
frequencies,
contribute to the radiation density at the time of BBN thereby
increasing the expansion rate and thus the light-element
abundances.  Measurements of light-element abundances
limit the number of additional relativistic species at BBN to
the equivalent of 1.4 neutrino degrees of freedom
\cite{Cyburt:2004yc}, which translates to a limit to a current
CGWB energy density $\Omega_{\mathrm{gw}} h^2 \lesssim 7.8\times
10^{-6}$.

The frequency range $\sim10^{-16} - 10^{-10}$ Hz remains largely
unconstrained.  An upper limit $\Omega_{\mathrm{gw}}h^2 \lesssim
0.1$ can be placed in this frequency range from QSO astrometry
\cite{Pyne:1995iy,Gwinn:1996gv}.
It has been proposed that future measurements of anisotropy in
the global rate of change of observed redshifts might someday
get down to $\Omega_{\mathrm{gw}}h^2 \sim 10^{-5}$ \cite{Seto:2005tq}.

Here we note that recent measurements of the angular
power spectrum of the cosmic microwave background (CMB)
that constrain the nonrelativistic-matter density $\Omega_m h^2$ to
roughly 10\% \cite{CMB} 
are, to a first approximation,
constraints to the radiation energy density at
the time of CMB decoupling; the constraint corresponds to a
limit of a few extra neutrino degrees of freedom.  From this,
we infer that the CMB provides
a limit to $\Omega_{\mathrm{gw}}h^2$ that may be competitive
with that from BBN, but extends to the lower frequencies,
$\sim10^{-15}$ Hz, corresponding to wavelengths comparable to
the comoving horizon at CMB decoupling \footnote{This estimate
was provided in Fig.~2 of Ref. \protect{\cite{Smith:2005mm}}.}.
This limit therefore improves upon previous constraints over the
frequency interval $10^{-15}-10^{-10}$ Hz by four orders of magnitude.

More precisely, the CGWB behaves as a
free-streaming gas of massless particles, just like massless
neutrinos, and therefore affect the growth of
density perturbations in ways in addition to their effect on the
expansion rate at decoupling.  If the CGWB
energy-density perturbations are adiabatic (i.e., have the same
density distribution as other relativistic species), then the effects of
the CGWB on the CMB/LSS are indistinguishable from those
due to massless neutrinos.  In this case, CMB/LSS constraints to
the number of massless neutrino species \cite{Pierpaoli:2003kw} translate
directly to a constraint to the CGWB energy density.
If, however, the primordial
perturbations to the CGWB energy-density
perturbations are non-adiabatic, as might be expected if they
are produced by inflation, pre-big-bang models, ekpyrotic, or
phase transitions and/or cosmic turbulence (see, e.g.,
Ref. \cite{Backgrounds}), then the CMB/LSS effects of
the CGWB may differ from those of adiabatic massless neutrinos.

In this paper, we carry out a detailed analysis of current
constraints to the CGWB amplitude that come from current measurements
of the CMB power spectrum and matter power spectrum.  Our
calculations of the CMB and matter power spectra include
the effects of the CGWB on the expansion rate and on the growth of
perturbations, for both adiabatic and non-adiabatic initial
conditions for the CGWB.  We include current constraints from
the CMB, galaxy surveys, and the Lyman-$\alpha$ forest.  We then
forecast how these constraints may be improved with future CMB
measurements.

\begin{figure}[!h]
\centerline{\epsfig{file= 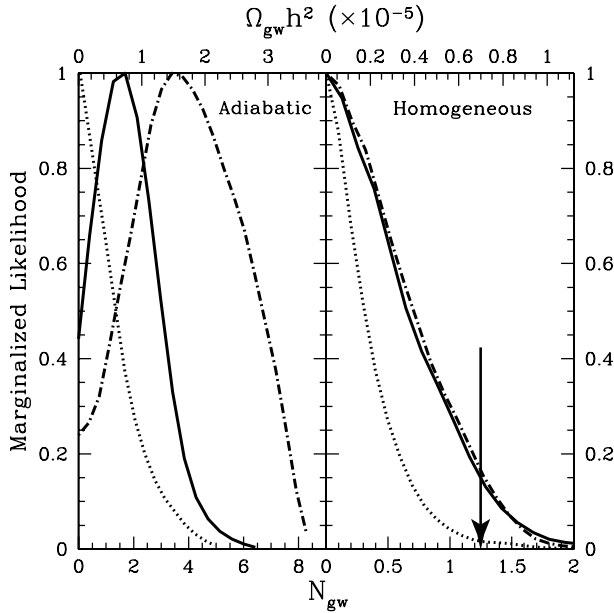, height=18pc,angle=0}}
\smallskip
\smallskip
\caption{{\it Adiabatic}: The marginalized (unnormalized) likelihoods for the CGWB
     energy density if perturbations to the CGWB
     density are adiabatic.  The dotted curve is the
     result obtained using only CMB data.  The thick solid curve
     includes galaxies as well as the Lyman-$\alpha$ forest.  In all of
     the aforementioned curves, the marginalization is over
     the nonrelativistic-matter density $\Omega_m h^2$, baryon
     density $\Omega_b h^2$, scalar spectral index $n_s$,
     power-spectrum amplitude $A_s$, the optical depth $\tau$ to
     the surface of last scatter, and the angle $\theta$
     subtended by the first acoustic peak (marginalization over
     $\theta$ essentially stands in for marginalization over the
     Hubble constant).  We hold the
     geometry fixed to flat, the number of neutrinos to
     $N_\nu=3.04$, and the neutrino masses fixed to zero.
     Finally, the dot-dash curve (to the right) shows current
     constraints from the CMB+galaxies+Ly$\alpha$ if we allow for and
     marginalize over nonzero neutrino masses as well.  The number of
     equivalent neutrino degrees of freedom ($N_{\mathrm{gw}}$) is shown on 
     the bottom axis. 
     {\it Homogeneous}:  same as the left panel, except for
     homogeneous initial conditions for the CGWB.  The arrow
     indicates the 95\% CL upper limit
     $\Omega_{\mathrm{gw}}h^2 \leq 6.9 \times 10^{-6}$ that we adopt as our central result.
     This is obtained from the analysis that includes
     current CMB+galaxy+Ly$\alpha$+free $m_{\nu}$.}
\label{fig:likelihood}
\end{figure}

We first consider the case when the CGWB has adiabatic initial
conditions.  In this case, the effects of the CGWB on the
expansion history and structure formation are identical to those
of massless neutrinos.  The analysis proceeds just as in
Ref. \cite{Pierpaoli:2003kw}.  We have updated this analysis to
include new small-scale CMB results, as well as constraints from
the Lyman-$\alpha$ forest.  The CMB results we use are from WMAP, 
ACBAR, CBI, 
VSA, and 
BOOMERanG, 
and we use the measurement of the galaxy power
spectrum from the 2dF Galaxy Redshift Survey and 
the Sloan Digital Sky Survey (SDSS), and 
the Lyman-$\alpha$ forest \cite{cosmomc_data}.  We
implement the Lyman-$\alpha$ constraints following the method discussed in Ref. \cite{Viel:2004np}, 
with minor modifications that were suggested by the authors. 
To translate the constraint to the
number of extra neutrino degrees of freedom to a
CGWB energy density, we use the relation
$\Omega_{\mathrm{gw}}h^2 = 5.6 \times 10^{-6}$, the density
contributed by a single massless-neutrino species.  

Results for adiabatic initial conditions are
shown in Fig.~\ref{fig:likelihood}.  
A limit at 95\% CL of $\Omega_{\mathrm{gw}}h^2 \lesssim
3.9 \times 10^{-5}$
is obtained from a combination of current
CMB data, galaxy power spectrum, and
the Lyman-$\alpha$ forest, and under the assumption that the
number of neutrino degrees of freedom is $N_\nu=3.04$ and that
neutrino masses are free to vary.  Due to a slight discrepancy between the
matter power spectrum from the best-fit CMB model and that
measured in galaxy surveys and Lyman-$\alpha$ forest measurements, 
the addition of galaxy surveys and the Lyman-$\alpha$ forest
weakens the bound by roughly a factor of two.
A small CGWB component improves slightly the
CMB+galaxy+Ly$\alpha$ agreement (cf., the solid curve in Fig.~\ref{fig:likelihood}), 
although the difference between $N_{\rm gw}=0$
and $N_{\rm gw}=2$ is statistically insignificant.  Although not shown,
we find that the exclusion of the Lyman-$\alpha$ forest weakens the 
CMB+galaxy+Ly$\alpha$ bound only
slightly.  If neutrino masses are assumed to be undetermined,
then the CMB+galaxy+Ly$\alpha$ bound is shifted by approximately two neutrinos (cf., the
dot-dash curve in Fig.~\ref{fig:likelihood}), which indicates that there is
a degeneracy between the neutrino mass and the CGWB.
This same trend has been observed in Ref.~\cite{Crotty:2004gm}. Note that the bound is improved by 
roughly a factor of 4 if we include only current CMB data.

\begin{table}[tb]\footnotesize
\caption{\label{tab:specs}}
\begin{center}
{\sc CMB Experimental Specifications for Fisher Matrix\\}
\begin{tabular}{rcccccc} 
\\ \hline \hline
Experiment & $\theta_{\rm beam}$ &  $(w_T)^{-1/2}$ &
$(w_P)^{-1/2}$ & $f_{\mathrm{sky}}$ & $\Omega_{\mathrm{gw}}h^2$\\ \hline \\
Planck: & 7.1 &  42.2 & 80.5 & 0.8 & $1.4 \times 10^{-6}$\\ & 5.0 &  64.8 & 132.3
& -- & --\\
\\ CMBPol: & 4.0 &  1.0 & 1.4 & 0.8 & $5 \times 10^{-7}$\\ \\ \hline
\end{tabular}
\end{center}
NOTES.---
The beam width, $\theta_{\rm beam}$, (FWHM) is given in arcminutes.   Weights, $(w_{T,P})^{-1/2}$, are in
arcminutes $\mu$K.  The sky fraction is given by $f_{\mathrm{sky}}$. The sensitivities to $\Omega_{\mathrm{gw}}h^2$ are 95\% CL for homogeneous
initial conditions. 
\end{table}

If the CGWB is produced from quantum fluctuations to the 
spacetime metric during the same superluminal expansion that
produced primordial density perturbations (e.g., from inflation,
but also from pre-big-bang or ekpyrotic scenarios), then
primordial perturbations to the CGWB density should be
non-adiabatic.  In such scenarios, the particle species in the
primordial Universe are all produced by decay of the inflaton;
this is why inflation produces primordial adiabatic
perturbations: i.e., the fractional perturbation to the energy
density of all the particle species are the same.  However,
gravitational waves are produced during
inflation by quantum fluctuations in tensor perturbations to the
spacetime metric---{\it not} through decay of the inflaton.
The CGWB should therefore not have the same primordial
energy-density perturbations as the particle species; in fact,
in linear theory, there should be no primordial
perturbation to the CGWB amplitude.  We therefore re-do our likelihood analysis
assuming the CGWB has homogeneous initial conditions.
More precisely, we have chosen to set the initial CGWB density
perturbation to zero in the conformal Newtonian gauge.  With
this {\it ansatz}, the primordial curvature perturbation
vanishes in the limit that the CGWB energy density dominates, as
it should; the curvature perturbation approaches the standard
adiabatic perturbation in the limit that the CGWB vanishes, also
as it should.  In this limit (appropriate for our analysis),
self-consistency of the perturbation equations demand that
nonzero higher-order moments in the CGWB distribution function
are induced at early times.  Details are given in
Ref.~\cite{inpreparation}.

If the CGWB is initially homogeneous, then the initial
conditions for the CGWB perturbations differ from those for
massless neutrinos.  This will affect the growth of
perturbations, especially at large scales~\cite{inpreparation},
and the degeneracy between the CGWB and massless neutrinos is
thus broken.  The bound to the CGWB then turns out to be
stronger than in the adiabatic case.
Fig.~\ref{fig:likelihood} shows results for the likelihood for
$\Omega_{\mathrm{gw}}h^2$ for different combinations of current
data sets as well as forecasts for the likelihoods expected when
future CMB experiments are included.  If the CGWB is produced by
some mechanism that leaves its primordial density uncorrelated
with the curvature perturbation---e.g., inflation or perhaps
some post-inflation phase-transition mechanism---then this is
the result that should be applicable.  We adopt as our 95\%
CL upper bound, $\Omega_{\mathrm{gw}}h^2 \lesssim 6.9 \times 10^{-6}$,
for homogeneous CGWB initial conditions from the combination of
data from current CMB experiments, galaxy surveys, and the
Lyman-$\alpha$ forest and under the assumption that the
number of neutrino degrees of freedom is $N_\nu=3.04$ and that
neutrino masses are free to vary.  Note, again, that the bound would be
roughly twice as strong if we were to restrict ourselves only to
CMB data.  And again, although not shown, we find that the exclusion 
of the Lyman-$\alpha$ forest weakens the 
CMB+galaxy+Ly$\alpha$ bound only slightly.

\begin{figure}[!h]
\centerline{\epsfig{file= 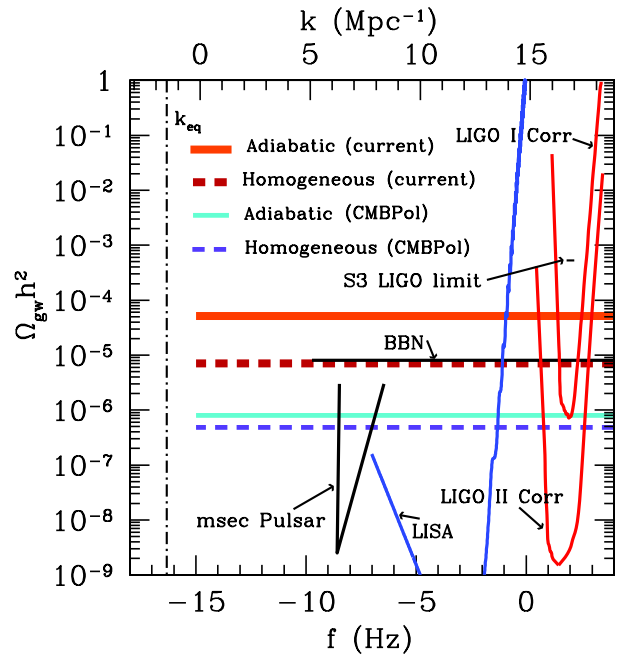, height=18pc,angle=0}}
\smallskip
\smallskip
\caption{The gravitational-wave density $\Omega_{\mathrm{gw}}
     h^2$ versus frequency.  The BBN constraint corresponds to a
     limit of 1.4 extra neutrino degrees of freedom.  We also
     show our constraints, from current CMB, galaxy, and
     Lyman-$\alpha$ data, for a CGWB with adiabatic primordial
     perturbations (``Adiabatic (current)'') and for 
     homogeneous initial conditions (``Homogeneous (current)''), as
     well as our forecasts for the sensitivities if current CMB
     data are replaced by data from CMBPol.  Also shown are the
     reaches of LIGO and LISA.  BBO
     (not shown) should go deeper, but primarily at
     frequencies $\sim1$ Hz.  Large-angle CMB fluctuations
     (also not shown) constrain $\Omega_{\mathrm{gw}}h^2
     \lesssim 10^{-14}$, but only at frequencies
     $\lesssim10^{-16}$ Hz.  The LIGO S3 upper limit is from Ref.~\cite{Abbott:2005ez}
     and the msec pulsar curve is from Refs.~\cite{Kaspi:1994hp, Lommen:2002je}.}
\label{fig:cgwb}
\end{figure}

Our central results are summarized in Fig.~\ref{fig:cgwb}, which
shows $\Omega_{\mathrm{gw}}h^2$ versus gravitational-wave
frequency.  Our new constraints are competitive with the BBN
constraint over the frequency range where both constraints
apply.  The precise value of the BBN constraint depends on the
precise constraint to the maximum number of neutrino degrees of
freedom allowed by BBN.  Some authors \cite{Burles:1999zt} claim a
limit $(N_\nu-3.04) \lesssim0.2$ (at 95\% CL), but more recent and
conservative estimates (that include new $^4$He measurements and
the CMB value for the baryon density) \cite{Cyburt:2004yc},
which we choose to adopt, place the limit
at $(N_\nu-3.04)\lesssim1.4$, comparable to the CMB/LSS bound we have
derived.  However, our new results apply four decades lower
in frequency, and provide the strongest constraint to the CGWB
amplitude over the frequency range $10^{-15}-10^{-10}$ Hz.  

To forecast the sensitivity of future CMB experiments to the CGWB, we have carried out a Fisher analysis that shows that when Planck
and CMBPol fly, the sensitivity
should be increased by a factor of roughly 10, while the BBN
constraint may continue to be limited by the same astrophysical
systematic uncertainties.  
See Table 1 for the experimental
specifications used in our Fisher analysis.  In our Fisher analysis we 
 included the improved CMB observations as well as the current galaxy 
 and Lyman-$\alpha$ constraints and allowed $m_{\nu}$ to vary with 
 $N_{\nu} = 3.04$.

We have not determined precisely the lower
end of the frequency range for which our bound applies.  In
order for the constraint to apply, the gravitational-wave
wavelength must be within the horizon at roughly the time of, or
slightly before, recombination.  Otherwise the waves do not
propagate as massless modes.  Analytic and numerical
integrations of the mode equations for gravitational waves in an
expanding Universe (e.g., Fig. 2 in
Ref.~\cite{Pritchard:2004qp}), indicate that the mode is
oscillating when $k\tau\simeq10$, where $k$ is the wavenumber
and $\tau$ the conformal time evaluated at decoupling.  This translates to a frequency
$\nu\simeq5\times10^{-17}$ Hz.  More realistically, the
gravitational wave will need to oscillate for a while before
recombination in order to have the effects we have considered
here.  We therefore tentatively estimate $10^{-15}$ Hz as the
lowest frequency for which our bound applies, although the
precise value may differ slightly.  We leave a more precise
calculation for future work.

There is also a slight correction if our bound is applied to a
scale-invariant spectrum.  In this case, the number of
gravitational-wave modes propagating as massless modes changes
with time, as more modes enter the horizon.  As a result, the
energy density does not scale with scale factor $a$ simply as
$a^{-4}$.  This, however, produces only a logarithmic
correction, which is within the theoretical error of the
treatment we have presented here.

Finally, we point out that the limit is probably not relevant
for scale-invariant spectra, such as those produced by
inflation.  Those are already constrained to be roughly eight
orders of magnitude lower in amplitude, at slightly lower
frequencies $\sim10^{-17}$ Hz, from large-angle fluctuations in
the CMB.  However, phase transitions or other exotic mechanisms
that produce a CGWB at frequencies $\gtrsim10^{-15}$ Hz will now
face this new constraint.

\smallskip
TLS was supported by an NSF Graduate Fellowship.  
EP is an ADVANCE fellow (NSF grant AST-0340648), 
also supported by NASA grant NAG5-11489.  This work was
supported in part by DoE DE-FG03-92-ER40701 and NASA
NNG05GF69G.  This work was supported in part by the NSF through the 
TeraGrid resources provided by NCSA and SDSC under TeraGrid grant 
AST050005T.

\end{document}